\documentclass{article}
\usepackage{graphicx}
\graphicspath{%
    {converted_graphics/}
    {C:/Dropbox/1-kgl-top/Papers/2016/CylindricalGrowth/}
}
\begin{document}

\begin{center}
{\LARGE Measurements of Cylindrical Ice Crystal Growth Limited by}\vskip6pt

{\LARGE \ Combined Particle and Heat Diffusion}\vskip16pt

{\Large Kenneth G. Libbrecht}\vskip8pt

{\large Department of Physics, California Institute of Technology}\vskip1pt

{\large Pasadena, California 91125}

\vskip18pt

\hrule\vskip1pt \hrule\vskip14pt
\end{center}

\textbf{Abstract.} We present measurements of the growth of long columnar
ice crystals from water vapor over a broad range of temperatures and
supersaturation levels in air. Starting with thin, c-axis ice needle
crystals, we observed their subsequent growth behavior in a vapor diffusion
chamber, extracting the initial radial growth velocities of the needles
under controlled conditions. Approximating the hexagonal needle crystals as
infinitely long cylinders, we created an analytical growth model that
includes effects from particle diffusion of water molecules through the
surrounding air along with the diffusion of heat generated by
solidification. With only minimal adjustment of model parameters, we
obtained excellent agreement with our experimental data. To our knowledge,
this is the first time that the combined effects from particle and heat
diffusion have been measured in ice growth from water vapor. This analysis
further provides an accurate method for calibration of the water-vapor
supersaturation levels in experimental growth chambers.

\section{Introduction}

\begin{figure}[htb] 
  \centering
  \includegraphics[height=4in,keepaspectratio]{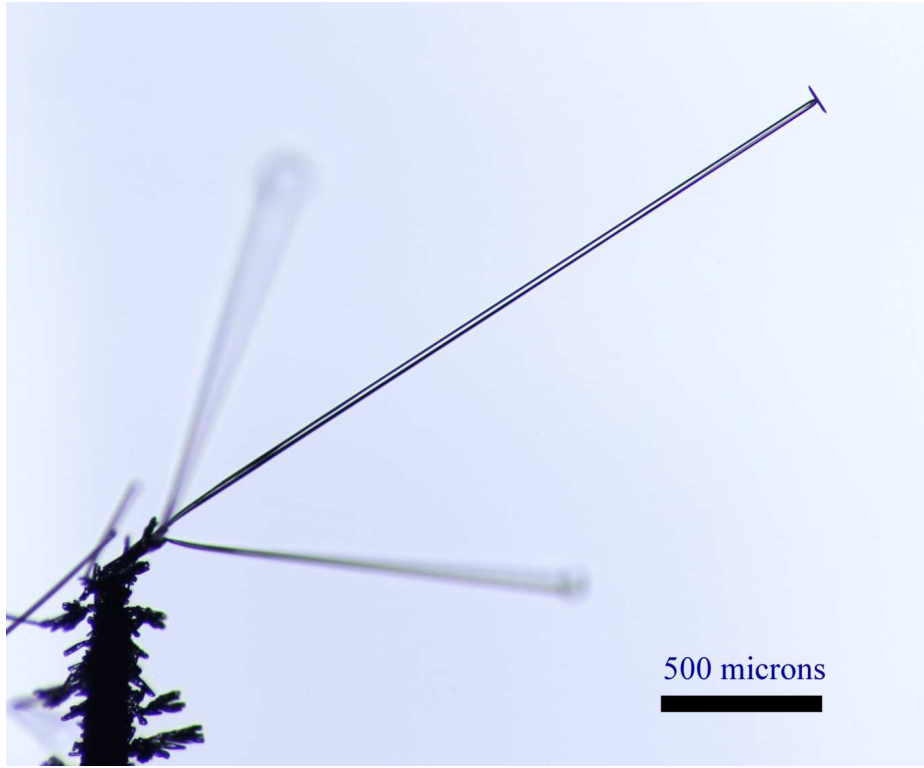}
  \caption{This photograph shows a typical
c-axis ice needle crystal growing at -15 C. The supporting wire substrate is
seen in the lower left of the photo, covered with a large number of frost
crystals. Several thin, c-axis ice needles grew out from the wire tip, and
one was brought into focus with the entire needle in the image plane. After
95 seconds of growth at a water-vapor supersaturation of $\protect\sigma %
_{center}\approx 11$ percent (in this particular example) a thin ice plate
can be seen growing on the tip of the ice needle, here seen in side view.
The ice plate diameter, ice needle diameter below the plate, and the overall
needle length can be extracted from a set of similar calibrated images.}
  \label{exampl}
\end{figure}

We recently developed a novel dual diffusion chamber for observing the
growth of ice crystals from water vapor in air, which allows us to create
slender needle crystals and measure their subsequent growth behavior under
carefully controlled conditions. The experimental apparatus is described in
some detail in \cite{kgldual14}. Figure \ref{exampl} shows an example of a
thin, plate-like ice crystal growing on the end of a long ice needle at a
temperature of -15 C. Our overarching goal with these observations is to
develop a comprehensive model of ice crystal growth from water vapor that
can reproduce quantitative growth rates as well as growth morphologies over
a broad range of circumstances. Although ice crystal formation has been
studied extensively for many decades, our understanding of the physical
effects governing growth behaviors at different temperatures and
supersaturations is still rather poor \cite{nakaya54, mason63, lamb72,
kurodalac82, lacmann83, nelsonknight98, libbrechtreview05}. 

Determining the water vapor supersaturation in ice growth experiments done
in air has long been a challenge, and it remains a significance hindrance to
making accurate, quantitative ice growth measurements. While a small
thermistor probe can easily determine air temperatures with excellent
absolute accuracy and little perturbation of the surrounding environment,
water vapor probes (hygrometers) are typically bulky and quite limited in
absolute accuracy. Moreover, in a supersaturated environment, water vapor
condenses on solid surfaces, and the presence of unwanted ice surfaces can
greatly affect the supersaturation field in their vicinity. As a result, one
often resorts to modeling of the experimental chamber to determine the
supersaturation within.

The second diffusion chamber in our dual-chamber apparatus was carefully
designed to facilitate accurate modeling of the water vapor supersaturation 
\cite{kgldual14}. The top and bottom surfaces have constant, well controlled
temperatures, and the side walls were constructed to maintain a simple,
linear vertical temperature gradient throughout the chamber. All walls of
the chamber are coated with ice crystals during operation, thus providing
well-defined boundary conditions for constructing a heat and water vapor
diffusion model of the interior of the chamber.

\subsection{Supersaturation in the Diffusion Chamber}

The temperatures of the top and bottom of the second diffusion chamber were
defined by $T_{top,bottom}=T_{set}\pm \Delta T,$ so $T_{top}-T_{bottom}=2%
\Delta T$ (see \cite{kgldual14} for the chamber dimensions), and the
temperatures of the four walls were maintained at $T_{walls}(z)=T_{bottom}+2%
\Delta T(z-z_{bottom})/(z_{top}-z_{bottom}).$ Solving the heat diffusion
equation within the chamber then yields the air temperature $%
T_{air}(z)=T_{walls}(z),$ and in particular we have $%
T_{center}=(T_{top}+T_{bottom})/2$ at the center of the chamber. Moreover,
the temperature gradient inhibits convective air currents within the
chamber. A shutter on one wall of the chamber is opened briefly to allow the
transport of crystals into the chamber, but it is otherwise kept closed to
maintain a stable temperature profile within the chamber.

If we first imagine moving the side walls of the chamber out to infinity,
then we can use a plane-parallel approximation to estimate the water-vapor
supersaturation at the chamber center, where test crystals are positioned.
Solving the diffusion equation for water vapor density $c(z)$ yields a
linear function with a constant gradient $dc/dz$ and $%
c_{center}=(c_{top}+c_{bottom})/2$, yielding the supersaturation at the
center of the chamber 
\begin{eqnarray}
\sigma _{center} &=&\frac{c_{center}-c_{sat}(T_{center})}{c_{sat}(T_{center})%
} \\
&=&\frac{1}{2}\frac{c_{sat}(T_{top})-2c_{sat}(T_{center})+c_{sat}(T_{bottom})%
}{c_{sat}(T_{center})}  \nonumber
\end{eqnarray}%
where $c_{sat}(T)$ is the saturated vapor pressure above an ice surface.
This expression gives the exact value for $\sigma _{center}$ in the
plane-parallel approximation (ignoring small changes in the diffusion
constant with temperature).

For small $\Delta T$, we expand the above expression to obtain the simpler
expression%
\begin{eqnarray}
\sigma _{center} &\approx &\frac{1}{2}\frac{1}{c_{sat}(T_{center})}\frac{%
d^{2}c_{sat}}{dT^{2}}(T_{center})\left( \Delta T\right) ^{2}  \label{cdiff}
\\
&\approx &C_{diff}(T_{center})\left( \Delta T\right) ^{2}  \nonumber
\end{eqnarray}%
The function $C_{diff}(T)$ can be calculated using $c_{sat}(T)\sim \exp
(-6150/T_{K})$ to good accuracy, where $T_{K}$ is the temperature in Kelvin.
In practice, we have found that the quadratic expansion is usually accurate
enough for our purposes, as it differs from the exact expression for $\sigma
_{center}$ by less than a percent when $\Delta T<6$ C, and it is only a few
percent high when $\Delta T=10$ C.

\begin{figure}[htb] 
  \centering
  \includegraphics[height=2.5in,keepaspectratio]{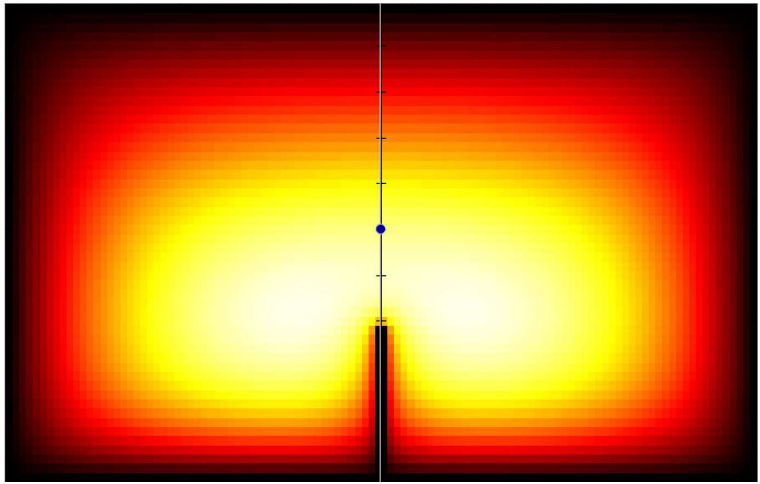}
  \caption{An example numerical model of
the second diffusion chamber described in \protect\cite{kgldual14}, showing
a contour plot of the water vapor supersaturation within the chamber. The
edges of the plot represent the outer walls of the chamber, and the
observation point is at the geometrical center of the chamber, marked here
with a round dot. Note that the supersaturation drops to zero at the chamber
walls (dark), and reaches its maximum value (white) below the center of the
chamber. In this particular model, the supersaturation also goes to zero
near an ice-covered central post that supports the test crystals.}
  \label{sigmamodel}
\end{figure}

To see how the chamber walls and the crystal support structure affected the
supersaturation field, we examined a range of computational models of the
diffusion chamber under different conditions, with one example shown in
Figure \ref{sigmamodel}. In these models we solved the dual-diffusion
problem (temperature and water-vapor density) numerically in three
dimensions. We found that, over a broad range of conditions near $%
T_{center}\approx -15$ C, the side walls reduced $\sigma _{center}$ by a
factor of approximately 0.8 compared to the plane-parallel approximation,
and an ice-covered central stem further reduced $\sigma _{center}$ by a
factor of approximately 0.9. Combined, our models indicated that these
effects lowered the supersaturation at the center by a constant geometrical
factor of $G_{{mod}}\approx 0.72$ compared with the plane-parallel
approximation, so that Equation \ref{cdiff} becomes%
\begin{equation}
\sigma _{center}(T_{center},\Delta T)\approx G_{{mod}}%
C_{diff}(T_{center})\left( \Delta T\right) ^{2}  \label{Cdiff2}
\end{equation}

Our models also indicated, however, that there remained significant
uncertainty in our ability to calculate $G_{{mod}}$, arising mostly
from difficulties in accurately modeling effects from the crystal support
structure and from a microscope objective placed inside the chamber about 90
mm away (horizontally) from the growing crystals.

\subsection{Modeling Cylindrical Crystal Growth}

As we began measuring crystal growth rates with this apparatus, we soon
realized that the radial growth of the ice needles could be used to
calibrate the supersaturation $\sigma _{center}$ as a function of $T_{center}
$ and $\Delta T.$ In essence, the prism surfaces of the columnar needles
serve as \textquotedblleft witness\textquotedblright\ surfaces, providing a
fairly accurate measure of the surrounding water vapor supersaturation. As
we will see below, this calibration works because the radial growth of the
needles is limited primarily by particle and heat diffusion, and is nearly
independent of the attachment coefficient $\alpha _{prism}$ at the needle
surface. Therefore we do not need to know $\alpha _{prism}$ with great
accuracy to calibrate $\sigma _{center}(T_{center,}\Delta T).$

To see this, consider the growth of an infinitely long cylindrical ice
crystal. Ignoring latent heat generation for the moment, we can solve the
particle diffusion equation in cylindrical coordinates to yield the general
solution $\sigma (r)=C_{1}+C_{2}\log (r),$ where $C_{1}$ and $C_{2}$ are
constants to be determined by the boundary conditions in the equation. Here
we have approximated the diffusion equation by Laplace's equation, which is
quite accurate in this situation, as the dimensionless Peclet number is much
less than unity \cite{libbrechtreview05}.

At the outer boundary of this model we assume a constant supersaturation $%
\sigma (R_{far})=\sigma _{far},$ where $R_{far}$ is the outer boundary of
the cylindrically symmetric diffusion field, and $\sigma _{far}$ is
essentially equal to $\sigma _{center}$ described above. (Note that $R_{far}$
cannot be set to infinity in cylindrical coordinates, as is commonly done in
spherical coordinates.) Equating $\sigma _{far}$ with $\sigma _{center}$
ignores the particle density gradient $dc/dz$ in the diffusion chamber,
which is justified by the observation that vertical asymmetries in crystal
growth rates are generally quite small.

At the inner boundary $R_{in}$, equal to the surface of the cylindrical
crystal, we write the radial growth velocity 
\begin{equation}
v=\frac{dR_{in}}{dt}=\frac{c_{sat}D}{c_{ice}}\frac{d\sigma }{dr}%
(R_{in})=\alpha _{prism}v_{kin}\sigma (R_{in})
\end{equation}%
where $D\approx 2\times 10^{-5}$ m/sec$^{2}$ is the particle diffusion
constant, $c_{ice}$ is the number density of ice, and $v_{kin}(T)$ is the
kinetic velocity defined in \cite{libbrechtreview05}. Including these
boundary conditions in the solution for $\sigma (r)$ then gives%
\begin{equation}
v=\frac{\alpha _{prism}\alpha _{diffcyl}}{\alpha _{prism}+\alpha _{diffcyl}}%
v_{kin}\sigma _{far}
\end{equation}%
where%
\begin{equation}
\alpha _{diffcyl}=\frac{1}{B}\frac{X_{0}}{R_{in}},  \label{alphadiff}
\end{equation}%
with $B=\log (R_{far}/R_{in})$ and $X_{0}=c_{sat}D/c_{ice}v_{kin}\approx
0.145$ $\mu $m. This cylindrical solution is similar to the spherical case
presented in \cite{libbrechtreview05}.

Using typical numbers (as we will see below) of $R_{in}=5$ $\mu $m and $%
R_{far}=2$ cm, we obtain the rather small value%
\[
\alpha _{diffcyl}\approx 0.0035
\]%
Comparing this $\alpha _{diffcyl}$ with the $\alpha _{prism}$ measurements
presented in \cite{kglalphas13}, we find that $\alpha _{diffcyl}\ll \alpha
_{prism}$ in most circumstances, allowing us to write%
\begin{equation}
v\approx \alpha _{diffcyl}v_{kin}\sigma _{far}  \label{vdiff}
\end{equation}%
and we see that this growth velocity is independent of $\alpha _{prism}$ as
long as $\alpha _{diffcyl}\ll \alpha _{prism}.$ Combining Equations \ref%
{Cdiff2}, \ref{alphadiff}, and \ref{vdiff} then yields the radial growth
rate of the cylinder%
\begin{equation}
v(R_{in})\approx \frac{G_{{mod}}}{B}\frac{X_{0}}{R_{in}}%
v_{kin}C_{diff}(T)\left( \Delta T\right) ^{2}  \label{v5-1}
\end{equation}

This equation gives us a good prediction for $v(R_{in}),$ as all the
parameters are rather tightly constrained except for $G_{{mod}}$ (which
is determined roughly by our modeling of the experimental chamber, as
described above) and $B=\log (R_{far}/R_{in}).$ However, Equation \ref{v5-1}
only applies in the absence of crystal heating from solidification, which
produces a significant perturbation of $v(R_{in})$, so we next examine
thermal considerations in our cylindrical crystal model.

Heating occurs because the growth of the cylindrical crystal releases a
latent heat per unit length of%
\[
\frac{dQ}{dLdt}=2\pi \lambda \rho vR_{in}
\]%
where $\lambda =2.8\times 10^{6}$ J/kg is the latent heat for the
vapor/solid transition and $\rho =917$ kg/m$^{3}$ is the ice density. This
generated heat must then be removed via conduction through the air
surrounding the crystal (ignoring convective air currents). Solving the heat
diffusion equation in cylindrical coordinates is similar to solving the
particle diffusion equation described above, and doing so yields a
temperature rise of the crystal (relative to the air temperature at $%
r=R_{far}$) given by%
\begin{equation}
\delta T=\frac{B\lambda \rho vR_{in}}{\kappa }  \label{heat}
\end{equation}%
where $\kappa =0.025$ W m$^{-1}$ K$^{-1}$ is the thermal conductivity of
air. The temperature rise increases the equilibrium vapor pressure of the
ice to%
\[
c(R_{in})\approx c_{sat}(T_{far})\left[ 1+\eta \delta T\right] 
\]%
where $\eta =d\log \left( c_{sat}\right) /dT$, and a bit of algebra reveals
that this reduces the growth rate to (see \cite{libbrechtreview05}) 
\begin{equation}
v(R_{in})\approx \frac{1}{1+\chi _{0}}\frac{G_{{mod}}}{B}\frac{X_{0}}{%
R_{in}}v_{kin}C_{diff}\left( \Delta T\right) ^{2}  \label{v5-2}
\end{equation}%
where%
\[
\chi _{0}=\frac{\eta D\lambda \rho }{\kappa }\frac{c_{sat}}{c_{ice}}
\]%
\qquad 

This result is an extension of Equation \ref{v5-1}, and $v\left(
R_{in},T,\Delta T\right) $ in Equation \ref{v5-2} provides our final
theoretical prediction for the growth rates of cylindrical crystals in our
diffusion chamber. Our next step is to compare these predicted growth rates
with experimental data.

\begin{figure}[htb] 
  \centering
  \includegraphics[height=4.0in,keepaspectratio]{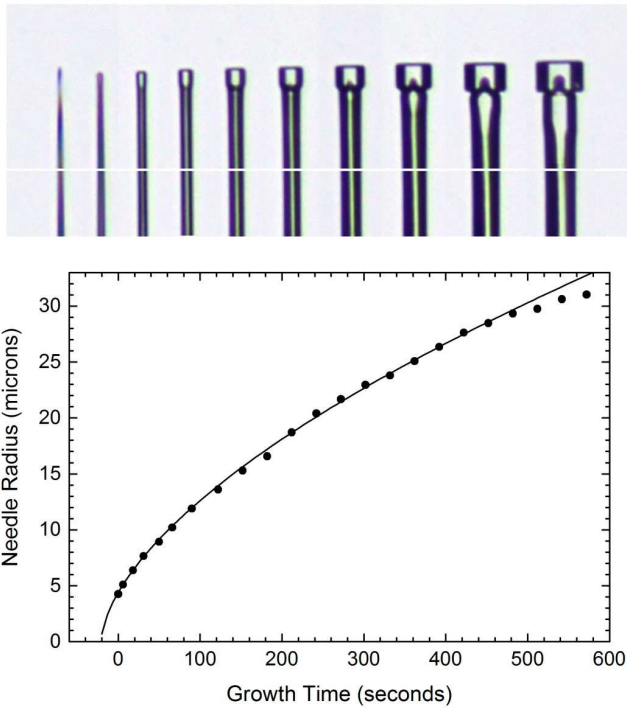}
  \caption{The top panel shows a composite
image of the tip of a single ice needle crystal as it grew. The graph below
it shows the needle radius at a location 100 $\protect\mu $m below the
needle tip, as indicated by the white horizontal line in the composite
image. The images correspond to the first ten points in the graph. The
needle grew in our diffusion chamber at a center temperature of $%
T_{center}=-2$ C with $\Delta T=7$ C.}
  \label{Rneedle}
\end{figure}

\section{Comparison with Crystal Growth Measurements}

For all the crystal growth data presented here, we first measured the needle
radius as a function of time $R_{needle}(t),$ at a location $L=100$ $\mu $m
below the needle tip, starting from images similar to that shown in Figure %
\ref{exampl}. Figure \ref{Rneedle} shows one example crystal grown at a
temperature of $T=-2$ C and $\Delta T=(T_{top}-T_{bottom})/2=7$ C. This
choice of $L$ was something of a compromise, being close enough to the
needle tip to be relevant for subsequent observations of ice structures at
the tip, while far enough below the tip that these same structures did not
greatly influence $R_{needle}(t)$ for small $R_{needle}.$

The optical microscope used to photograph the crystal had a resolving power
of 2.5 $\mu $m, and the image pixels measured 0.85 $\mu $m. Our diameter
resolution was therefore about $\pm 2$ $\mu $m, giving radial measurements
that were accurate to about $\pm 1$ $\mu $m. We then fit the $R_{needle}(t)$
data to a smooth curve to determine $v=dR_{needle}/dt$ at a time when $%
R_{needle}=5$ $\mu $m, as shown in Figure \ref{Rneedle}. We chose the
smallest practical $R_{needle}$ for which we could accurately measure $%
dR_{needle}/dt,$ because the tip structures more greatly perturbed the
cylinder growth at later times, when the tip structures (especially plates)
were larger in size. We did observe some variation in the measured $%
dR_{needle}/dt$ with changing $L$, with different tip structures, and
between different crystals grown in ostensibly the same conditions. But
these variations were at roughly the $\pm 20$ percent level, so they did not
alter our analysis greatly.

\begin{figure}[htbp] 
  \centering
  \includegraphics[height=6.7in,keepaspectratio]{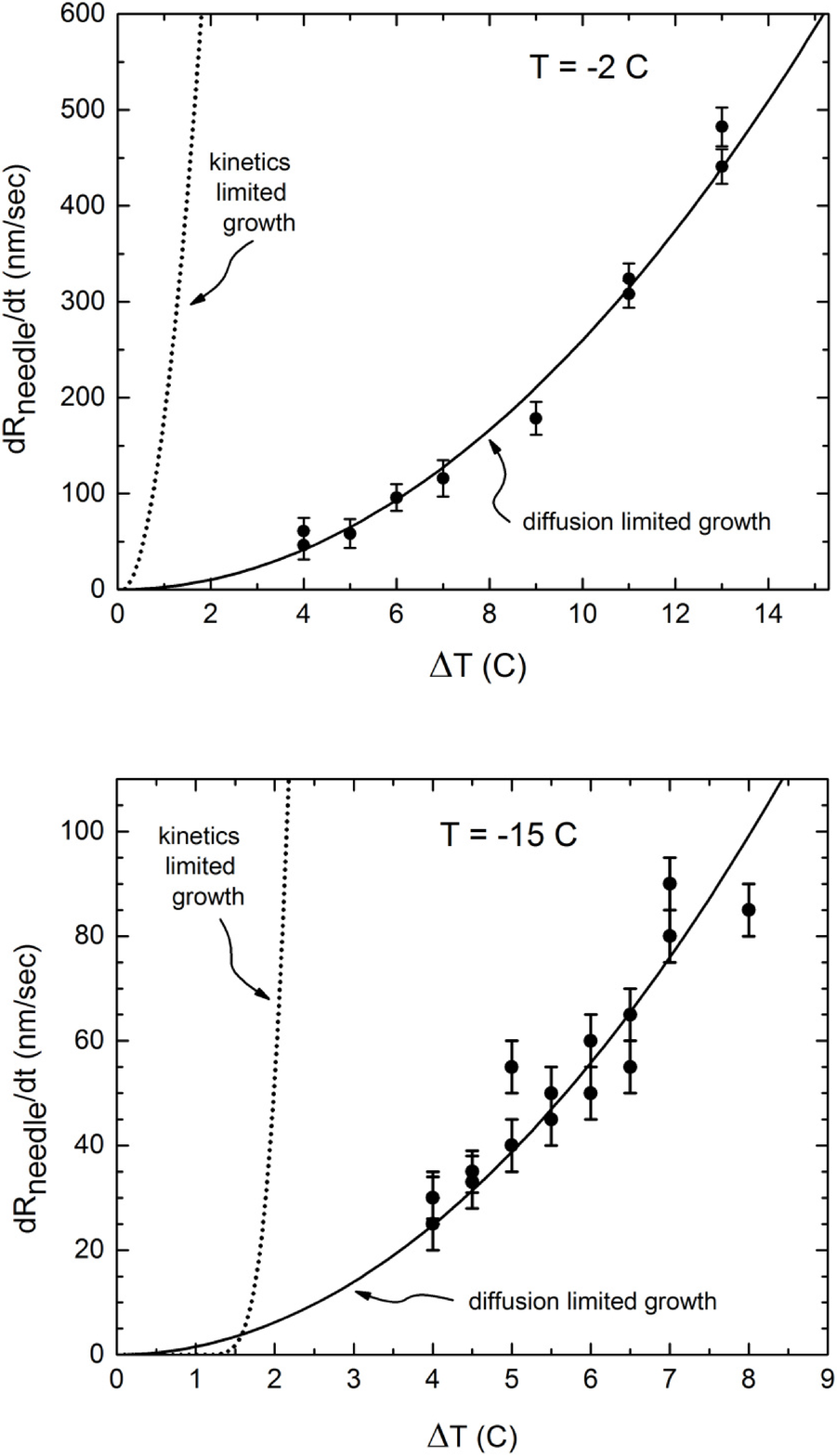}
  \caption{The upper graph shows
measurements of $v_{5}=dR_{needle}/dt$ when $R_{needle}=5$ $\protect\mu $m,
from data taken at a temperature of $T=-2$ C, as described in the text. The
lower graph shows similar data taken at $T=-15$ C. Both data sets were fit
to quadratic functions $v_{5}=A_{5}\Delta T^{2},$ shown as solid lines. This
functional form is expected if the growth is predominantly limited by
diffusion, and not by attachment kinetics. The dotted lines in both graphs
show models of what the velocities would be if the growth were limited
instead by attachment kinetics.}
  \label{RawData}
\end{figure}

\begin{figure}[htb] 
  \centering
  \includegraphics[height=4in,keepaspectratio]{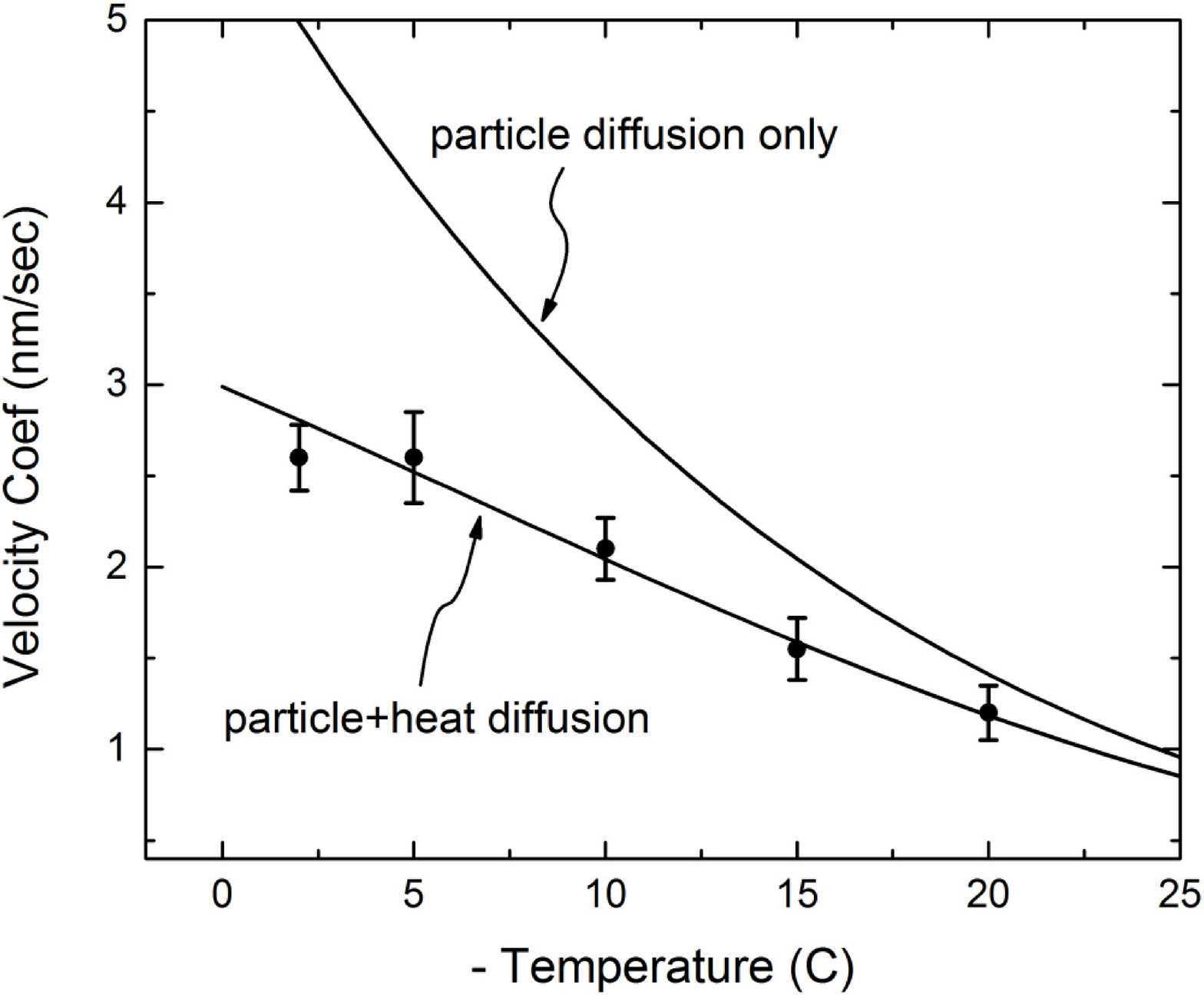}
  \caption{The points show the measured
velocity fit coefficient $A_{5}(T)$ as a function of the crystal growth
temperature. For comparison we show calculated $A_{5}(T)$ for particle
diffusion only (upper curve, Equation \protect\ref{v5-1}) and for combined
particle diffusion and heat diffusion (lower curve, Equation \protect\ref%
{v5-2}). Both curves used $G_{{mod}}=0.72$ and $R_{far}=2$ cm, after
adjusting $R_{far}$ so the lower curve best fit the observational data.}
  \label{velcoef}
\end{figure}

Figure \ref{RawData} shows the resulting radial growth velocity $%
v_{5}(T,\Delta T)=v(R_{needle}=5$ $\mu $m) measured at two representative
growth temperatures as a function of $\Delta T.$ These data, along with
similar data at other temperatures, were well fit with simple quadratic
functions $v_{5}(T,\Delta T)=A_{5}(T)\Delta T^{2}.$ The measured fit
coefficients $A_{5}(T)$ were then compared with calculated $A_{5}(T)$ from
Equation \ref{v5-2}, and the results are summarized in Figure \ref{velcoef}.
The theory curves used $G_{{mod}}=0.72$, and $R_{far}$ was adjusted to
fit the particle+heat diffusion curve to the data, yielding a best fit $%
R_{far}=2$ cm. As can be seen in Figure \ref{velcoef}, our data are in
excellent agreement with the expected particle+heat diffusion prediction
over the entire temperature range tested, with a physically reasonable
choice for $R_{far}.$

From this comparison between theory and experiment, we can extract a
prediction for $\sigma _{far}(T,\Delta T),$ essentially equal to $\sigma
_{center},$ the supersaturation at the center of the diffusion chamber in
the absence of any test crystals. Parameterizing this as $\sigma
_{far}(T,\Delta T)=A_{\sigma }(T)\Delta T^{2},$ the lower theory curve in
Figure \ref{velcoef} becomes the $A_{\sigma }(T)$ curve shown in Figure \ref%
{sigcoef}. The calculated $\sigma _{far}(T,\Delta T)=A_{\sigma }(T)\Delta
T^{2},$ using $R_{far}=2$ cm extracted from the data, then replaces Equation %
\ref{Cdiff2} as our best estimate of the supersaturation at the center of
our diffusion chamber, now calibrated using experimental data.

\begin{figure}[htb] 
  \centering
  \includegraphics[width=4.13in,height=3.2in,keepaspectratio]{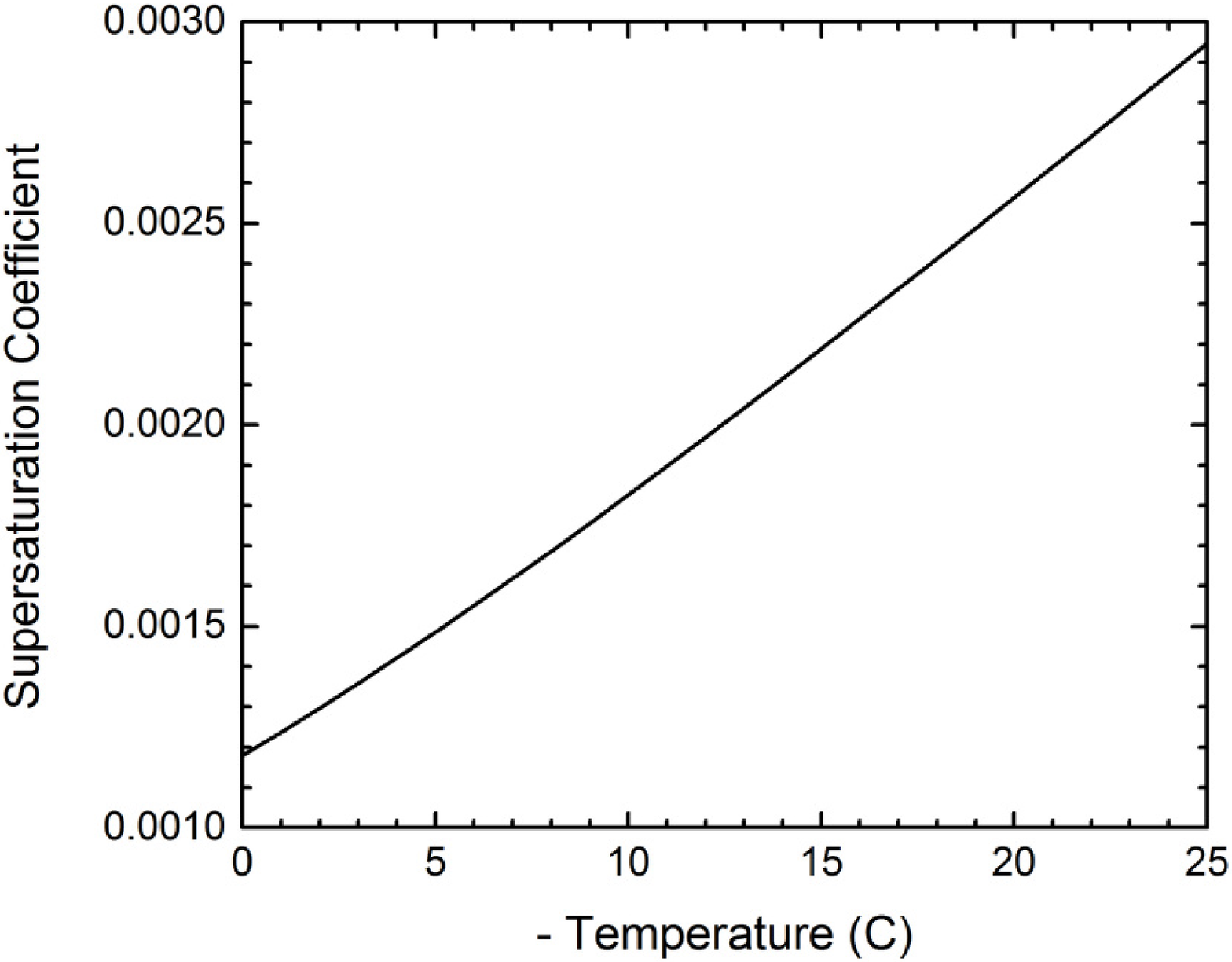}
  \caption{The final supersaturation coefficient $%
A_{\protect\sigma }(T),$ giving the effective $\protect\sigma _{far}=A_{%
\protect\sigma }(T)\Delta T^{2}$ for our experimental apparatus. The
cylindrical growth data thus calibrate the supersaturation in the growth
chamber, corrected for crystal heating, which is valuable for further
modeling of more complex growth morphologies.}
  \label{sigcoef}
\end{figure}

In summary, we have modeled the early growth of ice needle crystals in a
vapor diffusion chamber using a cylindrically symmetric approach that
approximates the needles as infinitely long cylinders. The largely
analytical model (with the geometrical correction factor $G_{{mod}}$
provided by numerical simulations) then yielded Equation \ref{v5-2}, which
gives the radial growth velocity $v(R_{in})$ as a function of $T,$ $\Delta T,
$ and other experimental parameters. Comparing this predicted $v(R_{in})$
with measurements at $R_{in}=5$ $\mu $m, we found excellent agreement using
a sensible value of $R_{far}=2$ cm for the outer boundary in the model. 

The data clearly indicate that both heat diffusion and particle diffusion
limit the crystal growth rates, as theory predicts. To our knowledge, this
is the first time that ice growth experiments have achieved sufficient
absolute accuracy to verify this basic theoretical prediction. Having a
reliable understanding of the supersaturation and resulting crystal growth
behavior for this simple cylindrical geometry is a major step forward in
producing accurate, quantitative measurements and models of more complex ice
crystal growth behaviors.

\bibliography{C:/Dropbox/1-kgl-top/Papers/1Bibliography/kglbiblio3}
\bibliographystyle{unsrt}

\end{document}